\documentclass[12pt]{article}
\begin{document}
\begin{center} {\large \bf Correlated Tunnelling of
Two Electrons through a Barrier in Quantum Wires}
\end{center}

S.A.Avdonin$^{1)}$, L.A.Dmitrieva$^{2)}$, Yu.A.Kuperin$^{3)}$, G.E.Rudin$^{3)}$

\vskip0.2cm
{\small \begin{center}
1) Department of Mathematical Sciences,\\
University of Alaska Fairbanks,
Alaska 99775-6660, USA\\
ffsaa@uaf.edu
\vskip0.2cm
2) Division of Mathematical and Computational Physics,\\
St.Petersburg State University,
198504 St.Petersburg, Russia\\
mila@JK1454.spb.edu

\vskip0.2cm
3) Laboratory of Complex Systems Theory,\\
St.Petersburg State University,
St.Petersburg 198504, Russia\\
kuperin@JK1454.spb.edu
\end{center}}

\vskip0.3cm
\begin{abstract}
We study the tunnelling trough a potential barrier of the system of two quantum
correlated particles.
The system is considered in one dimension.
The interaction with the barrier and between particles is approximated by $\delta$-potentials.
Assuming that the particles have the same masses, we reduce the problem to the set of independent planar scattering problems corresponding to two-body flux symmetry with respect to the barrier.
In order to solve the problems we apply the method of
Sommerfeld-Maluzhinets integral transformation, which requires
the solution of a set of functional equations. We show that the
sub-problem which is antisymmetric with respect to the center of mass allows the solution in frames of the Bethe ansatz.
So we give the exact solution of this problem.
For the subproblem which is symmetric we describe the class of
solutions for the functional equations and give an approach for
obtaining the solution.
\end{abstract}

\section{Introduction.}

The properties of low-dimensional semiconductor structures
have attracted much attention in the past
few years both experimentally and theoretically. Being the
quantum objects by nature the 1D and 2D nanostructures can be
described by an appropriate Schr\"odinger equation with the
suitable boundary conditions. There exit a huge literature on
the subject however the most results obtained deal with the
solution of 1-body or 2-body Schr\"odinger equation and with the
physical quantities that can be described in terms of these
solutions. To our knowledge there are only few results concerning
the 3-body systems in nanostructures while there are a lot of
physical phenomena which need the solution of 3-body quantum
problem. Let us mention only two of these phenomena:
\noindent
1) Observation of anomalies in the optical spectra of
quantum wells have been attributed to the formation of
negatively-charged excitons \cite{A0} . Such complexes can arise
when exciton $E$ captures one electron to form a
negatively-charged exciton $E+e \to E^-$ \cite{A1}. In ref.
\cite{A1} the properties of such 3-body complexes in 1D quantum
wires have been studied by means of exact solutions of 1D
3-body Schr\"odinger equation with the Calogero type interaction
\cite{A2}.
\noindent
2) Since the discovery of the vanishing transmissivity of a
tunnel barrier in a 1D quantum wire due to the electron-electron
interaction \cite{A3} , new interest has emerged in the
transport properties of 1D electron systems. Indeed, the
influence of the electron correlations shows up strikingly in
non-linear current-voltage relations which are investigated
experimentally in narrow semiconducting wires \cite{A4}.

From the mathematical point of view the type 2) of physical
phenomena can be described by 3-body Schr\"odinger equation in
one dimension if one treats the local electron interaction via
$\delta$-like potential and the quantum barrier (well) as the
third infinitely heavy particle which interacts with the
electrons by $\delta$-function potential as well. Solving the
appropriate 3-body 1D scattering problem one can calculate in
terms of scattering amplitudes such properties of tunneling
barrier in quantum wire as conductance, current-voltage
characteristics, etc. This scattering problem can be solved
either numerically or analytically i.e. exact.
The present paper is devoted to the exact solution of the
problem mentioned above by means of the reduction of the
original 3-body Schr\"odinger equation to some Riemann-Hilbert
problem. Using the symmetries of the system we split the
latter into two subproblems which are either symmetric with
respect to the center of mass or antisymmetric. For the
antisymmetric case we construct an exact solution in terms of
the Bethe anzatz. For the sub-problem which is symmetric we
suggest a new approach which allows to describe the functional
classes of the solutions and reduce the problem to some
generalized Dubrovin's equations.

The paper is organized as follows. In Sec.2 we briefly describe
the mathematical model for the correlated tunneling of two
electrons through a barrier. In Sec.3 by means of
Sommerfeld-Maluzhinetz transformation we construct the matrix
functional equations which have to be solved. In Sec.4 we give
an exact solution of antisymmetric sub-problem in terms of the
Bethe anzatz. Sec.5 is devoted to the study of the Riemann-Hilbert
problem for the symmetric subproblem. Sec.6 contains some
concluding remarks.

\section{The Model }

The model we study here has been suggested by K. Lipszyc \cite{LI} in 1980.
However, regardless the simple way the problem is posed, no closed-form solution is known until now.
Here we shall come closer to the exact solution of the complete problem and present the solution for one of two sub-problems it is reduced to.
Since the problem was carefully described in \cite{LI}, we shall allow us to abandon the details and present here just the topics essential for understanding of the following considerations (terms, notations, etc.).

We consider the system of two quantum particles with the same mass in one dimension interacting via delta-function potential with the strength $\gamma$. In addition, each particle is interacting in the same way with the fixed delta-function barrier with the strength $\Gamma$.
Let us denote the coordinates of the particles with respect to the barrier by $r_1$ and $r_2$, their masses by $m_1$ and $m_2$, choose the units so that $m_1 = m_2 = \hbar /2 = 1$.
The  Sr\"odinger equation for the wave function $u$ of the system can be written in the form:
\begin{equation}
[ - \Delta - \Gamma \delta ( r_1) - \Gamma \delta ( r_2)  -  \gamma \delta ( r_1 - r_2) ]
u ( r_1, r_2)  = \lambda u ( r_1, r_2).
\end{equation}
The correct mathematical formulation of this equation and the corresponding scattering
problem is given in terms of extension theory
\cite{PA, KMP} and was described for similar problems in \cite{KRR, KU, RU}.
Here we shall come directly to the equivalent boundary value problem. To do this,
let us introduce the polar coordinate system for
$(r_1, r_2) = (r \cos(\theta), r \sin (\theta))$, denote the function $u$ in
polar coordinates by $\tilde{u} $
and define
$$
[f](r,\phi) := f(r,\phi + 0) - f(r,\phi - 0).
$$
In this terms, the boundary value problem we shall study looks as \cite{DKR1}:
\begin{equation}
\label{BP1}
\begin{array}{c}
-\Delta \tilde{u}=\lambda \tilde{u} \\
\left.
\left[ \frac \partial {\partial _\phi } \tilde{u}  \right] ( r,\phi )=
- r \Gamma \tilde{u} ( r, \phi ) \right| _{\phi =0, \pi/2, \pi, 3\pi/2} \\
\left.
\left[ \frac \partial {\partial _\phi } \tilde{u}  \right] ( r,\phi )=
- 2 r  \gamma \tilde{u} ( r, \phi ) \right| _{\phi =\pi/4, 5\pi/4} \\
\end{array}
\end{equation}

Taking into account the symmetry of the system we split the solution of (\ref{BP1}) in two parts, symmetric and anti-symmetric with respect to the particle interchange:
\begin{equation}
\label{SPL1}
\begin{array}{c}
u (r_1, r_2) = u_-(r_1, r_2) + u_+(r_1, r_2)  \\
u_-(r_1,r_2) = - u_-(r_2,r_1) = (u (r_1, r_2) - u(r_2, r_1))/2\\
u_+(r_1,r_2) = + u_+(r_2,r_1) = (u (r_1, r_2) + u(r_2, r_1))/2
\end{array}
\end{equation}
The solution of (\ref{BP1}) for $u_-$ is trivial since the role of the second boundary condition (the interaction between particles) vanishes being replaced by the Dirichlet boundary condition $u(x, x) = \tilde{u}(r,\pi/4) = 0$. The expression for $u_-$ can be found as:
\begin{equation}
u_-(r_1,r_2) = f_1(r_1)  f_2(r_2) - f_1(r_2) f_2(r_1)
\end{equation}
where $f_{1,2}$ are the solutions of the one particle  problem with $\delta$-interaction .

We shall concentrate on the solution of the boundary problem (\ref{BP1}) for $u_+$.
Due to the symmetry of $u_+$  the boundary value problem (\ref{BP1})  can be expressed simpler in the centre of mass frame: $x = (r_1 + r_2) / \sqrt 2; y = (r_1 - r_2)/ \sqrt 2$, where $x$ is the $\sqrt 2$ times coordinate of the centre of mass and $y$ is the distance between particles divided by $\sqrt 2$.
Unless otherwise stated we shall use this coordinate system for $u_+$, the function $\tilde{u}_+$ will represent the same function in polar coordinates. In this notations, one has:
\begin{equation}
\label{BP2}
\begin{array}{c}
-\Delta \tilde{u}_+=\lambda \tilde{u}_+ \\
\left.
\left[ \frac \partial {\partial _\phi } \tilde{u}_+  \right] ( r,\phi )=
- r \Gamma \tilde{u}_+ ( r, \phi ) \right| _{\phi =\pi/4,  3\pi/4} \\
\left.
\frac \partial {\partial _\phi } \tilde{u}_+   ( r,\phi )=
- r \gamma \tilde{u}_+ ( r, \phi ) \right| _{\phi =0, \pi} \\
\phi \in (0, \pi)
\end{array}
\end{equation}

The problem (\ref{BP2}) can be further simplified using its symmetry with respect to the transformation
$x \rightarrow - x$ ($\phi \rightarrow \pi - \phi$). Let us split the function $u_+$ accordingly:
\begin{equation}
\label{SPL2}
\begin{array}{c}
u_+ (x, y) = v_-(x, y) + v_+(x, y)  \\
v_-(x, y)= - v_-(-x,y) = (u_+ (x, y) - u_+(-x,y))/2\\
v_+(x, y) = + v_+(-x,y) = (u _+(x, y) + u_+(-x,y))/2.
\end{array}
\end{equation}
The representation (\ref{SPL2}) splits (\ref{BP2}) into two independent problems for $v_\pm$:
\begin{equation}
\label{BP3}
\begin{array}{c}
\left[ \frac \partial {\partial _\phi } v_\pm  \right] ( r,\pi/4 )=
- r \Gamma v_\pm ( r, \pi/4 )  \\
\frac \partial {\partial _\phi } v_\pm+   ( r,0 )=
- r \gamma v_\pm ( r, 0 ) \\
1)\ v_- (r, \pi/2) = 0; \ \ \ 2)\ \frac \partial {\partial _\phi } v_+ (r, \pi/2) = 0\\
\phi \in (0, \pi/2)
\end{array}
\end{equation}
These are the boundary value problems we shall deal with in the rest of this paper.
In order to be rigorous we must add that we shall investigate the corresponding scattering problems. That is we shall look for the solutions of (\ref{BP3}) being the combination of one incoming plane wave $e^{-i k r \cos (\phi - \phi_0)}$, $\phi_0 \in (0, \pi)$, a set of outgoing plane waves, $\phi_0 \not \in (0, \pi)$, and a function satisfying the standard scattering conditions at $r \rightarrow \infty$.
The detailed description of how the scattering problem should be posed can be found in
extensive literature on the subject \cite{DKR1, MG1, MH1, MH2, LI, KU, KMP, PA}.

As a conclusion to this section, let us briefly discuss the physical meaning of the symmetries we applied.  The best visualization can be achieved if we discuss the  meaning of the corresponding incoming plane waves.
The plane wave in the system (\ref{BP1}) corresponds to the
incoming, that is in which the particles are directed towards
each other and towards the barrier, flow of pairs (since we
consider the interaction in pairs, we should consider the pair
of particles as one) of particles in which the momenta of the
corresponding particles are the same. That is all particles of
type 1 have the momentum $k_1$ and all particles of type 2 have
the momentum $k_2$. We should mention that though the particles
are assumed to have the same masses and interact with the barrier in the same way, they can, in principle, differ by some other properties (say by spin).
The plane wave in the system (\ref{BP2}) corresponds to two fluxes which have the momenta or particles interchanged and have the amplitudes which are either the same or differ by $-1$. In the case of absolutely identical particles, this would correspond to their statistics.
In the last case, the plane wave in the system (\ref{BP3}) corresponds
to four fluxes coming from two different sides to the barrier, by two from
each side, and whose amplitudes are either the same or differ by $-1$.
From the physical point of view, this situation as separate seems to be
hardly experimentally realizable but may also give a better understanding
of the process. In this paper we give the exact solution of the scattering problem
(\ref{BP3}) with boundary condition 1) and approach the solution of the (\ref{BP3}) with
condition 2).

\section{Sommerfeld-Maluzhinetz Integral Transformation and The Functional Equations.}

To discuss the solutions for (\ref{BP3}), K.Lipszyc \cite{LI} used the method of Sommerfeld-Maluzhinetz (SM) Integral Transformation \cite{MA}.
The Maluzhinets type functional equations were derived, however the solution was said to require some new unknown functional class to be used. Thus the problem was closed until further progress in functional analysis.
Here we shall also use the SM transformation, however we shall study the
functional equations in the form of McGuire \cite{MG1,MH2}. This way of using
the SM transformation was effectively applied to several
scattering problems \cite{KU,RU,KRR}.
In our previous papers \cite{MSRud,DKR1} we proved several
general statements which described the functional classes needed
to solve some SM functional equations (the McGuire type) and the
way one can construct their solutions. Here we apply these
statements to the concrete problem in question.

Let us apply the SM transformation to the function $v(r,\phi)$ considered on two domains:
$\tilde{v} (r, \phi) = v(r, \phi), \phi \in (0, \pi/4)$ and $\hat{v}(r, \phi) = v(r,\phi + \pi/4), \phi \in (0, \pi/4)$.
\begin{equation}
\label{SMT1}
\begin{array}{c}
\tilde{v} ( r,\phi ) =\int_c e^{-ikr\cos \alpha }
\left(
g_1( \alpha +\phi ) +g_2( \alpha +\pi /4-\phi )
\right) d\alpha \\
\hat{v} ( r,\phi ) =\int_c e^{-ikr\cos \alpha }
\left(
g_3( \alpha +\phi ) +g_4( \alpha +\pi /4-\phi )
\right) d\alpha
\end{array}
\end{equation}
where $c$ stands for the standard Sommerfeld contour \cite{MA,KU,MSRud}.
Using the representation (\ref{SMT1}) and the Maluzhinets theorem \cite{MA} the boundary value problem (\ref{BP3}) can be transformed into a set of functional equations:
\begin{equation}
\label{FE1}
\begin{array}{c}
ik \sin(\alpha)
\left(
g_3( \alpha ) - g_4( \alpha +\pi /4)
- g_1( \alpha + \pi/4) + g_2( \alpha)
\right)
\equiv
- \Gamma \left(
g_3( \alpha ) + g_4( \alpha +\pi /4)
\right)  \\
g_3( \alpha ) + g_4( \alpha +\pi /4) \equiv
g_1( \alpha + \pi/4) + g_2( \alpha)\\
ik \sin(\alpha)
\left(
g_1( \alpha) - g_2( \alpha + \pi/4)
\right)
\equiv
- \gamma \left(
g_1( \alpha ) + g_2( \alpha +\pi /4)
\right)  \\
g_{3\pm}( \alpha + \pi/4 ) \mp g_{2\pm}( \alpha ) \equiv 0
\end{array}
\end{equation}
Here we add the $+$ or $-$ signs to the functions corresponding to $v_+$ and $v_-$ respectively.

If we introduce the notations ${\bf g}( \alpha ) \equiv \left(
g_1( \alpha) , g_4( \alpha), g_3( \alpha), g_2( \alpha) \right) ^t$
the functional equations (\ref{FE1})  can be rewritten as
\begin{equation}
\label{MFE1}
{\bf g}( \alpha +\pi /4) =
M( \alpha ) {\bf g}( \alpha )
\end{equation}
where the meromorphic matrix function $M$ is given by
\begin{equation}
\label{Mmat}
M_\pm( \alpha ) =\left(
\begin{array}{cccc}
0 & 0 &
\frac{2 ik\sin(\alpha)}{2 ik\sin(\alpha) - \Gamma} &
\frac{\Gamma}{2 ik\sin(\alpha) - \Gamma} \\
0 & 0 &
\frac{\Gamma}{2 ik\sin(\alpha) - \Gamma} &
\frac{2 ik\sin(\alpha)}{2 ik\sin(\alpha) - \Gamma}\\
0 & \pm 1 & 0 & 0\\
\frac{ik\sin(\alpha) + \gamma}{ik\sin(\alpha) - \gamma}& 0 & 0 & 0
\end{array}
\right) .
\end{equation}

To study the functional equations (\ref{MFE1})  we need some algebraic results obtained in our previous papers \cite{DKR1,MSRud} for the generalized class of similar functional equations.
Below we apply these results to the problem under consideration.
Namely, let $M(x)$ and $F(x)$ be
meromorphic matrices depending on the complex variable $x$ and such that for
some $n\in {\bf N}$
\begin{equation}
\label{PropMF}
M\left( x+n\right) =M\left( x\right) ;\quad F\left( x+n\right)
=F\left( x\right) .
\end{equation}
The functional equation (\ref{MFE1}) with the argument shift $\pi/4$ can be reduced to the equation with the argument shift $1$ and the matrix coefficient having the period $n$ instead of $2 \pi$.
The properties of matrix coefficient $M(\alpha )$ from (\ref{Mmat}) can be
reconstructed if one poses $n=8$, $x=4 \alpha /\pi $.
We have considered
the matrix finite-difference equation
\begin{equation}
\label{MFEN}
g\left( x+1\right) =M\left( x\right) g\left( x\right) +F\left(
x\right) .
\end{equation}
We introduced the notation for ordered product of some matrices
$\{M_i\}$, namely $\uparrow \prod_{i=l}^kM_i\equiv M_kM_{k-1}...M_l$ for $
k\ge l$ and $\uparrow \prod_{i=l}^kM_i=I$ for $k<l$.
In this terms the $n$-th iteration of
the equation (\ref{MFEN}) reads as
\begin{equation}
\label{MFEIt}g\left( x+n\right) =N\left( x\right) g\left( x\right) +\Xi
\left( x\right)
\end{equation}
where we denoted $N\left( x\right) \equiv \uparrow \prod_{i=0}^{n-1}M\left(
x+i\right) $ and $\Xi \left( x\right) \equiv \sum_{i=0}^{n-1}\uparrow
\prod_{j=i+1}^{n-1}M\left( x+j\right) F(x+i).$ Due to the definition of $N$
one has
\begin{equation}
\label{PropN1}N\left( x+n\right) =N\left( x\right) ;\quad \Xi \left(
x+n\right) =\Xi \left( x\right)
\end{equation}
\begin{equation}
\label{PropN2}N\left( x+1\right) =M\left( x\right) N\left( x\right)
M^{-1}\left( x\right) .
\end{equation}
So the eigenvalues of $N\left( x\right) $ are of the period $1$.
Then we considered two special cases: a) $N\left( x\right) \equiv I$ b) $\det \left(
N\left( x\right) -I\right) \not \equiv 0$ which covered the three body problem with
zero-range interactions and are suitable for the problem we consider.
We need the following statements \cite{DKR1,MSRud}:

{\bf Lemma 1} {\em The general solution of inhomogeneous functional equation
$\left( \ref{MFEIt}\right) $ corresponding to the cases a) and b) looks like
\begin{equation}
\label{itsol}
\begin{array}{ll}
a) & g\left( x\right) =\frac xn\Xi \left( x\right) +g_0\left( x\right) ; \\
b) & g\left( x\right) =\left( I-N\left( x\right) \right) ^{-1}\Xi \left(
x\right) +g_0\left( x\right) ,
\end{array}
\end{equation}
where $g_0\left( x\right) $ is a solution of homogeneous functional equation
\begin{equation}
\label{HIFE}g_0\left( x+n\right) =N\left( x\right) g_0\left( x\right)
\end{equation}
}

Using this statement we were able to proof the

{\bf Theorem 1} {\em The general solution of inhomogeneous func\-ti\-onal
equation $\left( \ref{MFEN}\right) $ for the cases a) and b) looks like
\begin{equation}
\label{SolMFENa}a)\quad
\begin{array}{l}
g\left( x\right) =\sum_{i=0}^{n-1}\uparrow \prod_{j=i+1}^{n-1}M\left(
x+j\right) \left(
\frac{x+i}n\right) F(x+i)+ \\ +\sum_{i=0}^{n-1}\uparrow
\prod_{j=i}^{n-1}M\left( x+j\right) g_0(x+i),
\end{array}
\end{equation}
\begin{equation}
\label{SolMFENb}b)\quad
\begin{array}{l}
g\left( x\right) =\left( I-N\left( x\right) \right) ^{-1}\Xi \left( x\right)
+ \\
+\sum_{i=0}^{n-1}\uparrow \prod_{j=i}^{n-1}M\left( x+j\right) g_0(x+i),
\end{array}
\end{equation}
where $g_0\left( x\right) $ \ is a solution of homogeneous functional
equation $\left( \ref{HIFE}\right) $. For the case $\det N\left( x\right)
\not \equiv 0$ the solution can be also expressed as
\begin{equation}
\label{secrepr}
\begin{array}{c}
g\left( x\right) =\sum_{i=0}^{n-1}\uparrow \prod_{j=i}^{n-1}M\left(
x+j\right) \left( I-N\left( x+i\right) \right) ^{-1}F(x+i)+ \\
+\sum_{i=0}^{n-1}\uparrow \prod_{j=i}^{n-1}M\left( x+j\right) g_0(x+i).
\end{array}
\end{equation}
}

The statements above show that the first step in the study of the equations (\ref{FE1})
should be the construction of the 8-th iteration of the matrix $M(\alpha)$.
In order to simplify the calculations we introduce the notations
\begin{equation}
\label{NOT1}
\begin{array}{c}
\Gamma_k := \Gamma / k\\
\gamma_k := 2 \gamma / k\\
z := e^{i \alpha}
\end{array}
\end{equation}
and represent $M$ as $M = \pmatrix{0 & Q/q \cr P/p & 0\cr}$ where
\begin{equation}
\label{NOT2}
\begin{array}{c}
P_\pm(z) = \pmatrix{
0 &\pm (z^2 - \gamma_k z - 1) \cr
z^2 + \gamma_k z - 1 & 0 \cr}\\
p(z) = z^2 - \gamma_k z - 1 \\
Q(z) = \pmatrix{
z^2 - 1 &  \Gamma_k z ) \cr
\Gamma_k z & z^2 - 1 \cr} \\
q(z) = z^2 - \Gamma_k z - 1.
\end{array}
\end{equation}
This allows us, after one iteration, to deal with the $2\times 2$ matrices instead of $4\times 4$ ones, and consider the equation ${\bf g} (\alpha + 2 \pi) = N(\alpha) {\bf g}(\alpha)$ as a set of two independent equations.

\section{Anti-symmetric case and the Bethe ansatz}

Iteration of the equations for $v_-$, i.e. the case which is center-mass anti-symmetric with respect to the barrier, shows a surprising result:
$$
N_-(\alpha) \equiv I.
$$
As it was mentioned in \cite{DKR1} the latter result means that the Bethe ansatz can be used for the exact solution of the problem.
One can construct this solution through either SM transformation (see e.g. \cite{KU}) or directly solving the finite linear system for the plane waves coefficients.
Here we give the result of these calculations.

We shall construct the solution  as the decomposition of plane waves. Let us fix the wave vector $(k_1,k_2)$ for one of the plane waves. Using the reflection rules we conclude that the complete set of the plane waves can be written as:
\begin{equation}
\label{PW1}
\begin{array}{lc}
\phi \in (0, \pi/4) & v_-(x,y) = \sum_{l=1,..,8}{A_l e^{i{\bf k}_l{\bf x}} } \\
\phi \in (\pi/4, \pi/2) & v_-(x,y) = \sum_{l=1,..,8} { B_l e^{i{\bf k}_l{\bf x}} }
\end{array}
\end{equation}
where
\begin{equation}
\label{PWK1}
\left\{ {\bf k}_l \right\}_{l=1,..,8} =
\left\{
\begin{array}{rrrl}
(&k_1,&k_2)\\
(&-k_1,&-k_2)\\
(&k_1,&-k_2)\\
(&-k_1,&k_2)\\
(&k_2,&k_1)\\
(&-k_2,&-k_1)\\
(&k_2,&-k_1)\\
(&-k_2,&k_1)
\end{array}
\right\}
\end{equation}
and ${\bf x} := (x,y)$.

Let $\Re k_1\le 0$, $\Re k_1\le 0$ and $\Re k_1 < \Re k_2$, that is the wave vectors $(k_1,k_2)$ and $(k_2,k_2)$ correspond to the incoming waves which represent two situations: first, the particles are coming from the same side of the barrier; second the particles are coming from different sides of the barrier. To get the scattering amplitudes one should solve the boundary value problem (\ref{BP3}) for
$A_1 = 1; B_5 = 0$ (first case) and $A_1 = 0; B_5 = 1$ (second case).
The results we obtained for the scattering amplitudes are given
by equations  (\ref{SA1}) and (\ref{SA2}).\\

At   $A_1 = 1;\,\,  B_5 = 0$  we have:
\begin{equation}
\label{SA1}
\begin{array}{l}
{{{A_2}}=
     {{\frac{2\,\gamma \,{{{k_1}}^2}\,
           \left( \gamma  + i\,{k_2} \right)  +
          2\,{{{k_1}}^3}\,
           \left( -i\,\gamma  + {k_2} \right)  +
          {k_1}\,\left( -i\,\gamma  + {k_2} \right) \,
           \left( {{\Gamma }^2} - 2\,{{{k_2}}^2} \right)
            - \gamma \,\left( \gamma  + i\,{k_2} \right)
             \,\left( {{\Gamma }^2} + 2\,{{{k_2}}^2}
              \right) }{\left( \gamma  - i\,{k_1} \right)
            \,\left( \gamma  - i\,{k_2} \right) \,
          \left( i\,\Gamma  + {\sqrt{2}}\,{k_1} -
            {\sqrt{2}}\,{k_2} \right) \,
          \left( i\,\Gamma  + {\sqrt{2}}\,{k_1} +
            {\sqrt{2}}\,{k_2} \right) }}}}
\\
   {{{A_3}}=
     {-{\frac{\gamma  + i\,{k_2}}{\gamma  - i\,{k_2}}}}}
\\
   {{{A_4}}=
     {{\frac{-2\,\gamma \,{{{k_1}}^2} +
          2\,i\,{{{k_1}}^3} +
          i\,{k_1}\,\left( {{\Gamma }^2} -
             2\,{{{k_2}}^2} \right)  +
          \gamma \,\left( {{\Gamma }^2} +
             2\,{{{k_2}}^2} \right) }{\left( \gamma  -
            i\,{k_1} \right) \,
          \left( i\,\Gamma  + {\sqrt{2}}\,{k_1} -
            {\sqrt{2}}\,{k_2} \right) \,
          \left( i\,\Gamma  + {\sqrt{2}}\,{k_1} +
            {\sqrt{2}}\,{k_2} \right) }}}}
\\
{{{A_5}}=
     {-{\frac{\Gamma }
         {\Gamma  - i\,{\sqrt{2}}\,
            \left( {k_1} - {k_2} \right) }}}}
\\
   {{{A_6}}=
     {-{\frac{\Gamma \,\left( \gamma  + i\,{k_1} \right)
             \,\left( \gamma  + i\,{k_2} \right) }{
           \left( \gamma  - i\,{k_1} \right) \,
           \left( \gamma  - i\,{k_2} \right) \,
           \left( \Gamma  - i\,{\sqrt{2}}\,{k_1} -
             i\,{\sqrt{2}}\,{k_2} \right) }}}}
\\
{{{A_7}}=
     {{\frac{\Gamma \,\left( \gamma  + i\,{k_1} \right) }
        {\left( \gamma  - i\,{k_1} \right) \,
          \left( \Gamma  - i\,{\sqrt{2}}\,{k_1} +
            i\,{\sqrt{2}}\,{k_2} \right) }}}}
\\
   {{{A_8}}=
     {{\frac{\Gamma \,\left( \gamma  + i\,{k_2} \right) }
        {\left( \gamma  - i\,{k_2} \right) \,
          \left( \Gamma  -
            i\,{\sqrt{2}}\,\left( {k_1} + {k_2} \right)
             \right) }}}}

\\
{{{B_1}}=
     {{\frac{i\,{\sqrt{2}}\,
            \left( {k_1} - {k_2} \right)}
         {\Gamma  - i\,{\sqrt{2}}\,
            \left( {k_1} - {k_2} \right) }}}}
\\
{{{B_2}}=
     {{\frac{{\sqrt{2}}\,
          \left( \gamma  + i\,{k_2} \right) \,
          \left( {k_1} + {k_2} \right) }{\left( \gamma  -
            i\,{k_2} \right) \,
          \left( i\,\Gamma  + {\sqrt{2}}\,{k_1} +
            {\sqrt{2}}\,{k_2} \right) }}}}
\\
{{{B_3}}=
     {-{\frac{{\sqrt{2}}\,
           \left( \gamma  + i\,{k_2} \right) \,
           \left( {k_1} + {k_2} \right) }{\left( \gamma
               - i\,{k_2} \right) \,
           \left( i\,\Gamma  + {\sqrt{2}}\,{k_1} +
             {\sqrt{2}}\,{k_2} \right) }}}}
\\
   {{{B_4}}=
     { {\frac{ i\,{\sqrt{2}}\,
            \left( {k_1} - {k_2} \right)}
         {\Gamma  - i\,{\sqrt{2}}\,
            \left( {k_1} - {k_2} \right) }}}}
\\
{{{B_6}}=
     {{\frac{2\,{\sqrt{2}}\,\Gamma \,{k_1}\,
          \left( -i\,\gamma  + {k_2} \right) }{
          \left( \gamma  - i\,{k_1} \right) \,
          \left( i\,\Gamma  + {\sqrt{2}}\,{k_1} -
            {\sqrt{2}}\,{k_2} \right) \,
          \left( i\,\Gamma  + {\sqrt{2}}\,{k_1} +
            {\sqrt{2}}\,{k_2} \right) }}}}
\\
   {{{B_7}}=
     {{\frac{2\,i\,{\sqrt{2}}\,\Gamma \,{k_1}\,
          \left( \gamma  + i\,{k_2} \right) }{\left(
             \gamma  - i\,{k_1} \right) \,
          \left( i\,\Gamma  + {\sqrt{2}}\,{k_1} -
            {\sqrt{2}}\,{k_2} \right) \,
          \left( i\,\Gamma  + {\sqrt{2}}\,{k_1} +
            {\sqrt{2}}\,{k_2} \right) }}}}
\\
{{{B_8}}= 0}
\end{array}
\end{equation}

At  $A_1 = 0;\,\, B_5 = 1$  we   have:
\begin{equation}
\label{SA2}
\begin{array}{l}

\\
{{{B_1}}=
     {-{\frac{\Gamma }
         {\Gamma  - i\,{\sqrt{2}}\,{k_1} +
           i\,{\sqrt{2}}\,{k_2}}}}}
\\
   {{{B_2}}=
     {{\frac{-i\,\Gamma }
        {i\,\Gamma  + {\sqrt{2}}\,{k_1} +
          {\sqrt{2}}\,{k_2}}}}}
\\
   {{{B_3}}=
     {{\frac{i\,\Gamma }
        {i\,\Gamma  + {\sqrt{2}}\,{k_1} +
          {\sqrt{2}}\,{k_2}}}}}
\\
   {{{B_4}}=
     {{\frac{\Gamma }
        {\Gamma  - i\,{\sqrt{2}}\,{k_1} +
          i\,{\sqrt{2}}\,{k_2}}}}}
\\
   {{{B_6}}=
     {{\frac{2\,\gamma \,{{{k_1}}^2} +
          2\,i\,{{{k_1}}^3} +
          i\,{k_1}\,\left( {{\Gamma }^2} -
             2\,{{{k_2}}^2} \right)  -
          \gamma \,\left( {{\Gamma }^2} +
             2\,{{{k_2}}^2} \right) }{\left( \gamma  -
            i\,{k_1} \right) \,
          \left( i\,\Gamma  + {\sqrt{2}}\,{k_1} -
            {\sqrt{2}}\,{k_2} \right) \,
          \left( i\,\Gamma  + {\sqrt{2}}\,{k_1} +
            {\sqrt{2}}\,{k_2} \right) }}}}
\\
   {{{B_7}}=
     {{\frac{-2\,\gamma \,{{{k_1}}^2} -
          2\,i\,{{{k_1}}^3} -
          i\,{k_1}\,\left( {{\Gamma }^2} -
             2\,{{{k_2}}^2} \right)  +
          \gamma \,\left( {{\Gamma }^2} +
             2\,{{{k_2}}^2} \right) }{\left( \gamma  -
            i\,{k_1} \right) \,
          \left( i\,\Gamma  + {\sqrt{2}}\,{k_1} -
            {\sqrt{2}}\,{k_2} \right) \,
          \left( i\,\Gamma  + {\sqrt{2}}\,{k_1} +
            {\sqrt{2}}\,{k_2} \right) }}}}
\\
   {{{B_8}}= {-1}}
\\
   {{{A_2}}=
     {{\frac{2\,{\sqrt{2}}\,\Gamma \,{k_1}\,
          \left( -i\,\gamma  + {k_2} \right) }{
          \left( \gamma  - i\,{k_1} \right) \,
          \left( i\,\Gamma  + {\sqrt{2}}\,{k_1} -
            {\sqrt{2}}\,{k_2} \right) \,
          \left( i\,\Gamma  + {\sqrt{2}}\,{k_1} +
            {\sqrt{2}}\,{k_2} \right) }}}}
\\
   {{{A_3}}= 0}
\\
   {{{A_4}}=
     {{\frac{2\,{\sqrt{2}}\,\Gamma \,{k_1}\,
          \left( i\,\gamma  + {k_2} \right) }{\left(
             \gamma  - i\,{k_1} \right) \,
          \left( i\,\Gamma  + {\sqrt{2}}\,{k_1} -
            {\sqrt{2}}\,{k_2} \right) \,
          \left( i\,\Gamma  + {\sqrt{2}}\,{k_1} +
            {\sqrt{2}}\,{k_2} \right) }}}}
\\
   {{{A_5}}=
     {{\frac{{\sqrt{2}}\,\left( {k_1} - {k_2} \right) }
        {i\,\Gamma  + {\sqrt{2}}\,{k_1} -
          {\sqrt{2}}\,{k_2}}}}}
\\
   {{{A_6}}=
     {{\frac{{\sqrt{2}}\,
          \left( \gamma  + i\,{k_1} \right) \,
          \left( {k_1} + {k_2} \right) }{\left( \gamma  -
            i\,{k_1} \right) \,
          \left( i\,\Gamma  + {\sqrt{2}}\,{k_1} +
            {\sqrt{2}}\,{k_2} \right) }}}}
\\
   {{{A_7}}=
     {-{\frac{{\sqrt{2}}\,
           \left( \gamma  + i\,{k_1} \right) \,
           \left( {k_1} - {k_2} \right) }{\left( \gamma
               - i\,{k_1} \right) \,
           \left( i\,\Gamma  + {\sqrt{2}}\,{k_1} -
             {\sqrt{2}}\,{k_2} \right) }}}}
\\
   {{{A_8}}=
     {-{\frac{{\sqrt{2}}\,\left( {k_1} + {k_2} \right) }
         {i\,\Gamma  + {\sqrt{2}}\,{k_1} +
           {\sqrt{2}}\,{k_2}}}}}
\end{array}
\end{equation}

We are not able now to discuss the physical meaning of the results obtained in (\ref{SA1}) and (\ref{SA2}) because we need the similar expressions for the $v_+$  as part of the scattering amplitude. This is what we plan to do separately.
However, some basic features of solutions obtained with
(\ref{SA1}) and (\ref{SA2}) can be already discussed. Here we
shall  pay attention to the scattering of the particles being initially bound in a cluster.
The incoming cluster  is represented by the incoming surface wave:
$$
e^{-i k x - \gamma y }; \lambda = k^2 - \gamma^2.
$$
Substituting this into (\ref{SA1}) one has at
$ {{A_1}}= 1$ and $ {{B_5}}= 0$
\begin{equation}
\label{SA3}
\begin{array}{l}
{{{B_1}}=
     {1 + {\frac{\Gamma }
         {i\,{\sqrt{2}}\,k + {\sqrt{2}}\,\gamma  -
           \Gamma }}}}
\\
   {{{B_2}}= 0}
\ \
   {{{B_3}}= 0}
\\
   {{{B_4}}=
     {-1 + {\frac{\Gamma }
         {-i\,{\sqrt{2}}\,
            \left( k - i\,\gamma  \right)  + \Gamma }}}}
\\
   {{{B_6}}= 0}
\ \
   {{{B_7}}= 0}
\ \
   {{{B_8}}= 0}
\\
{{{A_2}}= 0}
\ \
   {{{A_3}}= 0}
\ \
   {{{A_4}}=
     {-{\frac{\left( k - i\,\gamma  \right) \,
           \left( 2\,{k^2} + 4\,i\,k\,\gamma  -
             2\,{{\gamma }^2} + {{\Gamma }^2} \right) }
           {\left( k + i\,\gamma  \right) \,
           \left( {\sqrt{2}}\,k -
             i\,{\sqrt{2}}\,\gamma  + i\,\Gamma  \right)
             \left( {\sqrt{2}}\,k +
             i\,\left( {\sqrt{2}}\,\gamma  + \Gamma
                 \right)  \right) }}}}
\\
   {{{A_5}}=
     {{\frac{\Gamma }
        {i\,{\sqrt{2}}\,k + {\sqrt{2}}\,\gamma  - \Gamma }
        }}}
\\
   {{{A_6}}= 0}
\\
   {{{A_7}}=
     {{\frac{\left( i\,k + \gamma  \right) \,\Gamma }
        {\left( -i\,k + \gamma  \right) \,
          \left( -i\,{\sqrt{2}}\,k -
            {\sqrt{2}}\,\gamma  + \Gamma  \right) }}}}
\\
   {{{A_8}}= 0}

\end{array}
\end{equation}

Considering (\ref{SA1}) we should first mention that the system in question does not
allow rearrangements, i.e. as the coupled particles come to the barrier they scatter
back also coupled. No transitions like: 2 $\leftrightarrow$ 1+1; 2 $\leftrightarrow$ 1+1* is possible. Here * stands for the particle bound at the barrier. This exceptional, in general, but common for the Bethe ansatz situation is due to the symmetry restrictions we applied.
The coefficient $A_4$ which plays the role of the reflection coefficient depends effectively on the strength of both interaction with the barrier and the interaction between particles. However in order to study the influence of both interactions on the cluster conductance through the barrier one should combine the data obtained for $v_-$ with ones for $v_+$.
In the next section we shall start the solution of the scattering problem for $v_+$.

\section{Symmetric case and the Riemann-Hilbert Problem}

The symmetric case, i.e. the case which is center-mass symmetric with respect to the barrier,
leads to the following expression
\begin{equation}
\label{N2}
N(z) = \pmatrix{N_1(z)/p_1(z) & 0 \cr 0 & N_2(z)/p_2(z)}
\end{equation}
through
the $2\times2$ matrixes $N_1(z)$ and $N_2(z)$ with polynomial coefficients and polynomials $p_1(z)$ and $p_2(z)$.
We do not present here the exact expressions for $N_i(z)$ and $p_i(z)$ we obtained by using the computerized tool for analytical calculations ({\it Mathematica}).
Instead of this we shall adopt our general results \cite{DKR1,MSRud} to the case of the
functional equation with $2\times2$ rational matrix coefficient as it has been done in (\ref{N2}).
In this way we can not obtain the exact solution for $v_+$ as it was done for $v_-$, however we are able to study the properties of this solution in much details.
Namely, for the application to the solution of scattering problem one needs to find analytic solutions of functional equations $\left( \ref{MFEN}\right) $ with
given singularities, which would correspond to the incoming plane waves \cite{MA,MG1}.
In order to do this it is useful to find a generating
solution, i.e. a meromorphic matrix function having meromorphic inverse one
with finite number of singularities in the band $\Re x\in [0,1)$. Then the
solution with given properties can be obtained through the action of
generating solution on some periodic vector function which is simply a
rational function of exponentials. A generating solution for the functional
equations (\ref{MFEN}) can be obtained from the solution of the iterated equations
(\ref{MFEIt}) using the results of Theorem 1. So in the rest of the paper we
shall concentrate on the functional equations (\ref{MFEIt}) only.

We reformulate the equations (\ref{MFEIt}) to the form of the boundary
value problem by introducing the variable $z=e^{2\pi i\frac xn}$,
which coincide with the definition (\ref{NOT1}) in the case period $2 \pi$.

Due to $\left( \ref{PropN1}\right) $ the matrix function $N\left(
z\right)$ will be meromorphic, which is just obvious in our case, and the boundary values of solutions for
(\ref{MFEIt}) on the cut $z\in [0,+\infty )$ will satisfy
\begin{equation}
\label{MFERHpr}g\left( z-i0\right) =N\left( z\right) g\left( z+i0\right)
+\Xi \left( z\right) .
\end{equation}
This is the special case of the problem known as the matrix Riemann-Hilbert
problem .
We propose the way to solve this type of
boundary value problems and demonstrate the functional class of solutions.

Let us define ${\cal L} := [0,+\infty)$, $z_{l-} := 0$, $z_{l+} := +\infty$, $\ln _{{\cal L}}z :=\ln z$.
Let us also note that the requirement
$N\left(z_{l\pm }\right) =I $ is obviously fulfilled in our case.
Here we need the results obtained for the
{\bf Homogeneous Riemann-Hilbert Problem: }{\em Find the vector function }$
{\bf g}\left( z\right) |{\bf C\mapsto C}^n${\em analytic on }$\overline{{\bf %
C}}{\bf \backslash }\left\{ {\cal L}\right\} $, {\em extendable until the
both sides of } ${\cal L}$ {\em and such that its boundary values satisfy}
\begin{equation}
\label{RH1}\left. {\bf g}\left( z\right) \right| _{{\cal L}_{+}}=\left.
N\left( z\right) {\bf g}\left( z\right) \right| _{{\cal L}_{-}}.
\end{equation}
The solution of an inhomogeneous problem would be needed if we study some singular solutions for $v_+$ (compare with \cite{KU}).

In our consideration we use the matrix coefficient $N\left( z\right) $
in the diagonal (or Jordan) form.
Here we can omit the $n\times n$ case, which we briefly
considered in our previous paper \cite{DKR1}  and come directly to the simple diagonalization of a $2\times 2$ matrix $N$.
Below we shall use the symbol $N(z)$ to refer any of $N_1/p_1$ or $N_2/p_2$.
Since any $2\times 2$ matrix has just two eigenvalues, which are given by the values of the two-sheet analytic function
$$
\lambda \left( z\right) =\frac 12\left( {\rm Tr}N\left( z\right) +\sqrt{{\rm %
Tr}^2N\left( z\right) -4{\rm Det }N\left( z\right) }\right),
$$
 the procedure of finding the diagonal or Jordan form is direct.
Moreover, for the problem in question we can find explicitly the expressions for
\begin{equation}
\label{N1D}
\begin{array}{l}
{\rm Tr}^2 N_1(z) - 4 {\rm Det} N_1(z) =
\\
-128 z^4 \left( -1 + {z^4} \right)^2 \gamma^2\Gamma^2
\\
(
1 + z^{16} +
(z^4 + z^{12})( -\gamma^4 + \Gamma^2 ( 4 + \Gamma^2)- 4\gamma^2( 1 + 2\Gamma^2))-
\\
z^8 ( 4 \gamma^2 ( 2 + \Gamma^4) + 2 ( 1 + 4\Gamma^2 + \Gamma^4)  +
       \gamma^4 ( 2 + 4\Gamma^2 +\Gamma^4)))
\end{array}
\end{equation}
and
\begin{equation}
\label{N2D}
\begin{array}{l}
{\rm Tr}^2 N_2(z) - 4 {\rm Det} N_2(z) =
\\
128 z^4 \left( 1 + {z^4} \right) ^2 \gamma^2\Gamma^2
\\
( 1 + z^{16} +
(z^4 + z^{12}) ( \gamma^4 - \Gamma^2 ( 4 + \Gamma^2)  + 4\gamma^2( 1 + 2\Gamma^2))+
\\
z^8( 4\gamma^2 ( 2 + \Gamma^4)  + 2 ( 1 + 4\Gamma^2 + \Gamma^4)  +
      \gamma^4( 2 + 4\Gamma^2 + \Gamma^4))) .
\end{array}
\end{equation}
We conclude that we deal now with the case (1.b) of \cite{DKR1},
Let us mention that the eigenvalues
$\lambda _1\left( z\right) \ne $$\lambda _2\left( z\right) $
a.e. and have branching points.

The corresponding eigenprojections $\Lambda _i\left( z\right) $ can be expressed as
$\Lambda_i\left( z\right) =
{\bf v}_i\left( z\right) \times \widetilde{{\bf v}}_i^t\left( z\right)
$ where ${\bf v}_i\left( z\right) $ is the eigenvector
of $N\left( z\right) $ corresponding to the eigenvalue $\lambda _i\left(
z\right) $ and $\widetilde{{\bf v}}_i^t\left( z\right) $ is one for $N\left(
z\right) ^t$. These eigenvectors are
normalized as $\widetilde{{\bf v}}_i^t\left( z\right) \times {\bf v}_i\left(
z\right) =1$, so ${\rm Tr}\Lambda _i\left( z\right) =1$, $\Lambda _i^2\left(
z\right) =\Lambda _i\left( z\right) $ and $\Lambda _j\left( z\right) \Lambda
_i\left( z\right) =0$ for $i\ne j$.

Let us recall the notion of the generating solution we used \cite{DKR1}.
We call an analytic matrix function $X\left( z\right) $ {\em generating
solution \ } of the Riemann-Hilbert problem \footnote{%
Note that the generating solution for the functional equation $\left( \ref
{MFEIt}\right) $ can be obtained as $X\left( e^{2\pi i\frac xn}\right) $.} $%
\left( \ref{RH1}\right) $ if

\begin{enumerate}
\item  it satisfies the boundary value problem
\begin{equation}
\label{RH1m}\left. X\left( z\right) \right| _{{\cal L}_{+}}=\left. N\left(
z\right) X\left( z\right) \right| _{{\cal L}_{-}}.
\end{equation}

\item  $X\left( z\right) $ and $X^{-1}\left( z\right) $ are regular on ${\bf %
C/\{}{\cal L}\}$ except may be some finite number $s$ of points $\left\{
z_i\right\} _{i=1}^s$ where they poles of finite multiplicities.
\end{enumerate}

Suppose we know some generating solution of the problem $\left( \ref{RH1}
\right) $. One can show that for any regular solution ${\bf g}\left(
z\right) $ the function ${\bf g}_0\left( z\right) \equiv X^{-1}\left(
z\right) {\bf g}\left( z\right) $ should satisfy
$$
\left. {\bf g}_0\left( z\right) \right| _{{\cal L}_{+}}=\left. {\bf g}
_0\left( z\right) \right| _{{\cal L}_{-}}
$$
and so would be rational function with poles at the points $\left\{
z_i\right\} $ only and of multiplicity not higher then one of $X^{-1}\left(
z\right) $.
In general, any such function can be expanded in finite linear combination
of simplest fractions and the regularity conditions at the points $\left\{
z_i\right\} $ would lead to the homogeneous system of linear equations for
the coefficients in this decomposition. However, we shall
demonstration this fact elsewhere.

In \cite{DKR1} we proposed the way how to construct a generating
solution for the general case of meromorphic matrix $N(z)$. Here
we shall show how this can be applied to the specific problem in question. To find the
generating solution we make use of the fact that the matrix coefficient can be represented as
$$
N\left( z\right) =\lambda _{+}\left( z\right) \Lambda _{+}\left( z\right)
+\lambda _{-}\left( z\right) \Lambda _{-}\left( z\right)
$$
where the signs $\pm $ denote the values of functions on different sheets of
the Riemann surface.
In the considered case the structure of the Riemann surface is determined in (\ref{N1D}) or
(\ref{N2D}), i.e. by the algebraic curves

\begin{equation}
\label{K1}
\begin{array}{l}
K_1(z) =
\left(
1 + z^{16} +
(z^4 + z^{12})( -\gamma^4 + \Gamma^2 ( 4 + \Gamma^2)- 4\gamma^2( 1 + 2\Gamma^2))-
\right.
\\
\left.
z^8 ( 4 \gamma^2 ( 2 + \Gamma^4) + 2 ( 1 + 4\Gamma^2 + \Gamma^4)  +
       \gamma^4 ( 2 + 4\Gamma^2 +\Gamma^4))
\right)^{\frac{1}{2}}
\end{array}
\end{equation}
or
\begin{equation}
\label{K2}
\begin{array}{l}
K_2(z) =
\left(
1 + z^{16} +
(z^4 + z^{12}) ( \gamma^4 - \Gamma^2 ( 4 + \Gamma^2)  + 4\gamma^2( 1 + 2\Gamma^2))+
\right.
\\
\left.
z^8( 4\gamma^2 ( 2 + \Gamma^4)  + 2 ( 1 + 4\Gamma^2 + \Gamma^4)  +
      \gamma^4( 2 + 4\Gamma^2 + \Gamma^4))
\right)^{\frac{1}{2}}
\end{array}
\end{equation}
which have by 16 branching points $\left\{ h_i\right\} _{i=1}^a$ in all non-degenerated cases, i.e. $\Gamma\ne 0$, $\gamma \ne 0$. The points $h_i$ are given by simple but bulk expressions we shall leave outside this text.
Let us split the set $\left\{ h_i\right\} _{i=1}^{16}$ in two parts $\left\{
h_i\right\} _{i=1}^a=\left\{ u_i\right\} _{i=1}^8\cup \left\{ v_i\right\}
_{i=1}^8$ and fix the cuts $\delta _i$ which run between the points $u_i$
and $v_i$ so that they do not cross the half-axis $[0,\infty)$.
Because the construction below is the same for both functions $K_1(z)$ and $K_2(z)$ we shall omit
the indices and refer these functions as $K(z)$.
We denote by $K_{\pm }\left( z\right) $  two branches of the
function $K(z)$. The branches can be fixed by the condition
$$
\lim_{z\rightarrow \infty} K_+(z)/z^8 = 1.
$$
The corresponding branches of $\lambda \left( z\right) $ and $%
\Lambda \left( z\right) $ are denoted as $\lambda _{\pm }\left( z\right) $
and $\Lambda _{\pm }\left( z\right) $. If we denote the points of the
Riemann surface ${\cal R}$ of $K\left( z\right) $ by ${\bf z}\equiv \left(
z,j\right) $; $z\in {\bf C}$; $j=\pm $ these branches are defined as $\lambda
_{\pm }\left( z\right) \equiv \lambda \left( \left( z,\pm \right) \right)
;\quad \Lambda _{\pm }\left( z\right) \equiv \Lambda \left( \left( z,\pm
\right) \right) $.

Let us define the function $
g\left( {\bf z}\right) $ on ${\cal R}$
\begin{equation}
\label{SRHonRSsol}
g_0\left(
{\bf z}\right) =
\exp \left\{
\frac { K( {\bf z})} {2\pi i} \int_0^{+\infty}
\frac{\ln \lambda_+ (z^\prime) }
{ K_+\left(z^\prime\right)
\left(z^\prime-z\right) }
dz^\prime
\right\}
\end{equation}
We have to mention that (\ref{SRHonRSsol}) gives a new
representation of $g_0({\bf z})$ in comparing with $g_0({\bf
z})$ from \cite{DKR1}. This form is more compact and seems to be
more suitable for the actual calculations since it does not require the calculation of $\lambda(z)$ singularities.

The function $g_0\left( {\bf z}\right) $ is meromorphic on ${\cal R}%
\backslash \left\{ \infty \right\} $ and has the essential singularity at $%
z\rightarrow \infty $. As $z\rightarrow \infty $ the function $g_0\left(
{\bf z}\right) $ behaves as $e^{q\left( z\right) }$ where the coefficients
of the polynomial $q\left( z\right) $ of the degree 8.

This singularity can be canceled if one multiplies $g_0\left( {\bf z}\right)
$ by a product of the Baker-Ahiezer (BA) function $\psi \left( {\bf z}\right) $ of the Riemann surface ${\cal R}$ corresponding to the point $\infty $, the
polynomial $q\left( z\right) $ and some non-special divisor $D$ of degree $n$
\cite{Dubrovin}. Really, if one fixes $n$ points on ${\cal R}$ and allows the
function $\psi \left( {\bf z}\right) $ to have simple poles at this points,
i.e. specifies the divisor $D$, then the condition
$\psi \left( {\bf z} \right) \sim e^{-q\left( z\right) }$ at ${\bf z\rightarrow }\infty $ will
specify the function $\psi $ uniquely up to the constant factor. In order to
get the explicit expression for the BA function one chooses the canonical
sections $\left\{ a_i\right\} _{i=1}^8$ and $\left\{ b_i\right\} _{i=1}^8$
on the Riemann surface ${\cal R}$. Then one constructs the second kind
differential $\Omega $ with the main part $-dq\left( z\right) $ at $\infty $
and normalized by the conditions
$$
\oint_{a_i}\Omega =0.
$$
When one denotes by
$U=\left\{ U_1,..,U_m\right\} $ the vector of $b$ -periods of $\Omega $,
$$
U_i=\oint_{b_i}\Omega
$$
and chooses some point ${\bf z}_0$ on ${\cal R}$ the
explicit expression for $\psi $ looks like \cite{Dubrovin}
\begin{equation}
\label{BakAhF}\psi \left( {\bf z}\right) =\exp \left( \int_{{\bf z}_0}^{{\bf %
z}}\Omega \right) \frac{\theta \left( A\left( {\bf z}\right) -A\left(
D\right) +U-K\right) }{\theta \left( A\left( {\bf z}\right) -A\left(
D\right) -K\right) }
\end{equation}
where $K$ is the vector of Riemann constants, $A$ denotes the Abel transform
and $\theta $ is the Riemann theta-function of the surface ${\cal R}$.

Let us introduce the function $g\left( {\bf z}\right) =g_0\left( {\bf z}
\right) \psi \left( {\bf z}\right) $. The branches of $g\left( {\bf z}
\right) $ on the sheets ${\cal R}$ we shall denote by
$$
g_{+}\left( z\right) \equiv g\left( \left( z,+\right) \right) ;\quad
g_{-}\left( z\right) \equiv g\left( \left( z,-\right) \right) .
$$
In this notations the generating solution of the Riemann-Hilbert problem can
be found as
$$
X\left( z\right) =g_{+}\left( z\right) \Lambda _{+}\left( z\right)
+g_{-}\left( z\right) \Lambda _{-}\left( z\right) .
$$

The regularization of $g_0\left( {\bf z}\right) $ described above has some
shortcomings. Namely, the investigation of $X^{-1}\left( z\right) $ meets
the problem of theta-function zeroes and the formulae obtained
are not a good base for the numerical calculations. In the concluding section of
\cite{DKR1} we proposed another method to construct the BA function. The method leads
to the set of Dubrovin's type \cite{Dubrovin} ordinary differential equations.
We plan, in the further study on the subject,
to investigate the solutions of this equations for the particular case we study here.
However we should emphasize that the method of the solutions construction for (\ref{MFE1})
and their functional class is already studied.

\section{Conclusion}

In summary, we have studied a new mathematical model for
correlated tunneling of electrons through a barrier in 1D quantum
wires. For one subproblem the exact solution has been obtained
and for another one the reduction to the generalized Dubrovin's
equation has been constructed. The latter can be solved either
numerically or analytically. The future work will be devoted to
the exact integration of the Dubrovin's equations obtained and
to the calculation in terms of their solutions of the physical
characteristics of correlated tunneling in quantum wires.

Finally we note that the approach developed in this paper can be
easily applied to the phenomena of type 1) mentioned in
Introduction, i.e. to the study of exotic exciton complexes in
1D semiconducting nanostructures.

\section{Acknowledgments}

The authors are indebted to B.S.Pavlov for drawing attention to the problem.

\end{document}